\documentclass[conference]{IEEEtran}
\IEEEoverridecommandlockouts
\usepackage{cite}
\usepackage{amsmath,amssymb,amsfonts}
\usepackage{algorithmic}
\usepackage{graphicx}
\usepackage{textcomp}
\usepackage{xcolor}

\usepackage{breakcites}
\usepackage{indentfirst}
\usepackage{pgfgantt}
\usepackage{pdflscape}
\usepackage{float}
\usepackage{epsfig}
\usepackage{epstopdf}
\usepackage{multirow}

\usepackage[utf8]{inputenc}
\def\BibTeX{{\rm B\kern-.05em{\sc i\kern-.025em b}\kern-.08em
    T\kern-.1667em\lower.7ex\hbox{E}\kern-.125emX}}

\makeatletter
\newcommand{\newlineauthors}{%
  \end{@IEEEauthorhalign}\hfill\mbox{}\par
  \mbox{}\hfill\begin{@IEEEauthorhalign}
}
\makeatother    
    
\begin{document}

\title{Network-Aware AutoML Framework for Software-Defined Sensor Networks}

\author{\IEEEauthorblockN{Emre Horsanali}
\IEEEauthorblockA{\textit{Department of Computer Engineering}\\
\textit{Istanbul Technical University}\\
Istanbul, Turkey \\
Email: horsanalie18@itu.edu.tr}
\and
\IEEEauthorblockN{Yagmur Yigit}
\IEEEauthorblockA{\textit{Department of Computer Engineering}\\
\textit{Istanbul Technical University}\\
Istanbul, Turkey\\
Email: yigity20@itu.edu.tr}
\and
\IEEEauthorblockN{Gokhan Secinti}
\IEEEauthorblockA{\textit{Department of Computer Engineering}\\
\textit{Istanbul Technical University}\\
Istanbul, Turkey\\
Email: secinti@itu.edu.tr}
\newlineauthors
\IEEEauthorblockN{Aytac Karameseoglu}
\IEEEauthorblockA{\textit{BTS Group}\\
Istanbul, Turkey\\
Email: akarameseoglu@btsgrup.com}
\and
\IEEEauthorblockN{Berk Canberk}
\IEEEauthorblockA{\textit{Department of AI and Data Engineering}\\
\textit{Istanbul Technical University}\\
Istanbul, Turkey\\
Email: canberk@itu.edu.tr}
}

\maketitle

\begin{abstract}
As the current detection solutions of distributed denial of service attacks (DDoS)  need additional infrastructures to handle high aggregate data rates, they are not suitable for sensor networks or the internet of things. Besides, the security architecture of software-defined sensor networks needs to pay attention to the vulnerabilities of both software-defined networks and sensor networks.
In this paper, we propose a network-aware automated machine learning (AutoML) framework, which detects DDoS attacks in software-defined sensor networks. 
Our framework selects an ideal machine learning algorithm to detect DDoS attacks in network-constrained environments, using the metrics such as variable traffic load, heterogeneous traffic rate, and detection time while preventing over-fitting. 
Our contributions are two-fold: (i) we first investigate the trade-off between the efficiency of ML algorithms and network/traffic state in the scope of DDoS detection. (ii) we design and implement a software architecture containing open-source network tools, with the deployment of multiple ML algorithms. 
Lastly, we show that under the denial of service attacks, our framework ensures the traffic packets are still delivered within the network with additional delays.

\end{abstract}


\section{Introduction}
\label{sec:intro}
The use of wireless sensor networks is increasing everyday, providing a crucial backbone for future IoT applications spanning from military to smart cities. 
%
%
Furthermore, through network function virtualization and resource abstraction, -although, it is developed for wired infrastructures- software-defined network approach paves it way into wireless domain during the last decade; providing invaluable flexibility in network orchestration and accelerating research and development cycle.  
This revolutionary changes drastically shift how we monitor, control, communicate in any environment and  high level policies to be deployed in various domains, which otherwise would be unimaginable. 
These technologies combined offer new and unique attributes such as data and control plane distinction, centralized control, distributed data aggregation and processing with limited life-cycle of sensor nodes.  
With these new attributes in mind, the conventional network security methods/algorithms should be definitely revised to address and overcome the possible network vulnerabilities efficiently. 
Security of the network infrastructure is one of the top concerns in a heterogeneous environment where multiple sensors participate in a single platform such as software-defined sensor networks (SDSN) \cite{iot2019}. Network infrastructure and sensors in the network belong to different manufacturers, even if all infrastructure devices belong to the same manufacturer, this makes the management and control of the network difficult and increases the vulnerability to launch an attack on the entire network using a single compromised device/sensor \cite{survey44}. The limited computing power in sensor devices makes them vulnerable to being used as bots for various attacks such as DDoS \cite{IoT2017}, \cite{transSDN2018}. Sensor nodes in SDSN are highly prone to DDoS attacks due to the lack of resources to implement a local security solution to defend against these attacks \cite{challengeSDSN}. The existing detection solutions are not suitable to handle high aggregate data rates result from sensors since they need additional infrastructures for sensor networks or the internet of things \cite{trans20intro}. Since SDSN includes two domains, software-defined network and sensor network, the vulnerabilities in these domains also apply to SDSN. Therefore, the security architecture of SDSN needs to pay attention to itself, the controller, the sensor nodes, and the communication protocols \cite{challengeSDSN}, \cite{africon2017}.
In this paper, we aim to utilize machine learning (ML) based solution to mitigate DDoS attack in software-defined sensor networks.  
In this manner, rather than focusing on a single ML method to address DDoS attacks, we propose an AutoML framework with network-aware model selection heuristic. 

AutoML is one of the flourishing areas of booming AI/ML research, receiving a lot of attention lately. 
AutoML aims to automate the time consuming and costly processes based on hyperparameter optimization, combinatorial optimization, and transfer learning. Hyperparameter optimization is used to adjust the hyperparameters, combinatorial optimization is used to optimize performance by combining different features and transfer learning is used to accelerate model configuration by transferring learning data and parameters \cite{19MLAuto}.
Therefore, we suggest an AutoML framework to dynamically select an ideal ML algorithm to utilize for the task of detecting DDoS attacks in network-constrained environments such as SDSN.
Our framework benefits from three major network/system parameters such as variable traffic payloads, node power levels, and heterogeneous traffic speeds in order to form a network-awareness.
Using this viable information, our AutoML framework determines the most suitable algorithm, by taking network state and power-levels of nodes into consideration while preventing over-fitting. 

We list  main contributions of the paper as follows:
\begin{itemize}

    \item We exhaustively explore the trade-off between crucial network parameters and ML algorithms. In this manner, we use variable payload size and heterogeneous traffic models to model the network state and evaluate the accuracy and computation cost of multiple ML algorithms. 
    \item We proposing an AutoML framework selecting optimal ML algorithm not only by considering the accuracy of the model but also by taking sensor network challenges into account. Furthermore, we design and implement an open-source software stack architecture offering simple SDSN testbed with adaptable ML module.
\end{itemize}

The rest of the paper is organized as follows. 
Section~\ref{sec:preliminary} gives preliminary experiment and Section~\ref{sec:relWork} provides a survey of related studies in the literature. The network-aware AutoML framework and the performance evaluation are explained in Section~\ref{sec:autoML} and Section~\ref{sec:perfEval} respectively. We conclude the paper in Section~\ref{sec:conc}.

\section{Preliminary Experiments}
 
Before starting our implementation, we examine the ping round trip time graph from the beginning to the end of the DDoS attacks in a simple SDSN environment. Furthermore, we also examine how the system behaves during a DDoS attack when it is utilizing a greedy algorithm for detecting DDoS.

\begin{figure}[htbp]
\begin{minipage}[htbp]{0.49\linewidth}
    \includegraphics[width=\textwidth]{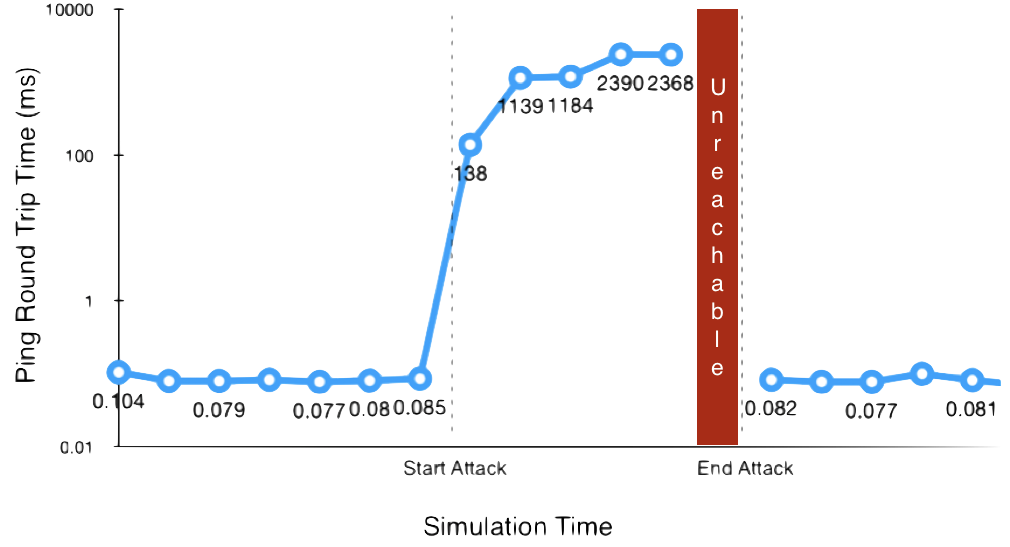}
    \caption{Round Trip Time: the simple SDSN behavior under DDoS attacks}
    \label{fig:5at}
\end{minipage}%
\hfill%
\begin{minipage}[htbp]{0.49\linewidth} 
    \includegraphics[width=\textwidth]{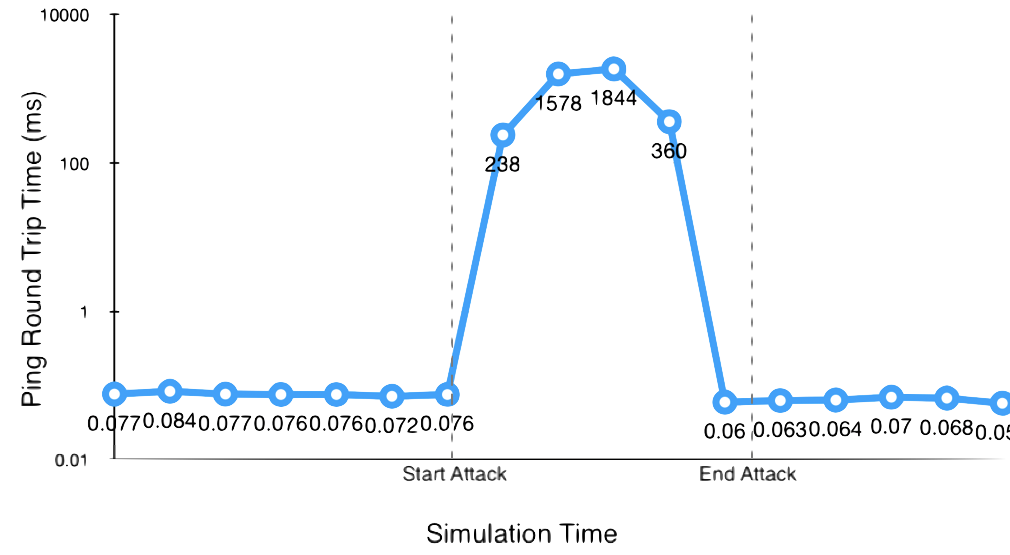}
    \caption{Round Trip Time: the simple SDSN behavior using the greedy algorithm under DDoS attacks}
    \label{fig:5bt}
\end{minipage}        
\end{figure}

Figure~\ref{fig:5at} depicts that the system is an unreachable state for a substantial amount of time when there is no implementation to detect and mitigate DDoS attacks.

As mentioned above, we also utilize a greedy algorithm to observe system behavior under DDoS attacks. The problem of relevant resources' unreachability during the attack is eliminated thanks to the greedy algorithm, but still, access time to the relevant resources is quite high for sensor networks, as can be seen in figure~\ref{fig:5bt}. Therefore, we decide to improve the round trip time graph using machine learning algorithms and propose our system architecture.

\section{Related Work}
\label{sec:relWork}
The SDN architecture has its unique requirements for security and there are number of works focused on SDN security in the literature. For instance, M. Ariman et al. proposed a traffic-aware routing algorithm add-on which manages the flows their QoS requirements and they implemented a real-time testbed for software-defined wireless networks (SDWN) by using Raspberry Pi \cite{Ariman}. 

To be more precise, we particularly surveyed the articles that focus on the DDoS attack detection solution in SDN using machine learning techniques in the literature. 
Some of these works proposed only detection solutions and the others proposed both detection \cite{det2} and mitigation solutions \cite{detmit2} in SDN. Liang Tan et al. proposed a framework for detection and mitigation of DDoS attacks using a combined machine learning algorithm based on K-Means and K-Nearest Neighbors (KNN) to detect suspicious flows by the detection trigger mechanism in the SDN environment. They implemented a trigger mechanism on the data plane to screen for abnormal flows in the network \cite{20Access}. DETPro framework which is an efficient and lightweight DDoS attack detection and mitigation system is proposed based on the decision tree method \cite{19ICC}. Shi Dong et al. suggested two methods to detect the DDoS attack in SDN, the first of these methods uses the degree of DDoS attack and the other uses the improved K-NN algorithm \cite{20AcKNN}. N. N. Tuan et al. proposed a TCP-SYN flood attack mitigation framework in SDN networks using KNN-based algorithm \cite{19ICTC}. In the other work, flexible modular architecture is introduced that allows the identification and mitigation of low rate DDoS attacks in SDN which trained using six machine learning models (i.e., J48, Random Tree, REP Tree, Random Forest, Multi-Layer Perceptron (MLP), and SVM) \cite{20LowRate}. O. Rahman et al. evaluated J48, Random Forest (RF), SVM, and K-NN algorithms to detect and block the DDoS attack in an SDN network and the results showed that J48 has better performance than others \cite{19Congress}.
K. Duran et al. proposed an automated attribute-based IP management module with the three engines and they succeeded in a more accurate anomaly detection method by decreasing the anomaly scoring rate compared to IPAMv1 \cite{kubra}.  
SVM, Naive Bayes, Decision Tree, and Logistic Regression techniques were evaluated to detect DDoS attacks in SDN and the results showed that SVM has better performance \cite{20Globe}. 
The other work, an AI-driven partial topology discovery approach is proposed to maintain a global MAC service which serves both physical and virtual connections in a broadband network \cite{9369564}.
Furthermore, R. Santos et al. provided the implementation of four machine learning algorithms (SVM, MLP, Decision Tree, and RF) to detect DDoS attacks in three different categories (flow-table attack, bandwidth attack, and controller attack) \cite{2019wiley}. Their results showed that the RF algorithm has the best accuracy and the Decision Tree algorithm has the best performance. 
E. Ak et al. proposed a two-scale fair sensitivity control mechanism to solve the unfair channel access problem followed by the increase in the hidden/exposed STAs problem using multilayer perceptron in the WLANs \cite{Ak}.

Security is still a pretty open topic that has not yet received adequate attention in SDSN. Most of the current work at SDSN focuses on the architectural framework, in part because the field is new \cite{challengeSDSN}. Some works in the literature focused on DDoS detection in software-defined sensor networks. G. A. N. Segura et al. proposed a lightweight and efficient DDoS attack detection approach using change point analysis; this approach is suitable for WSNs \cite{ICC2020}. In the other work, an SD-IoT based framework that provides security services to the IoT network was proposed developing a counter-based DDoS attack detection application based on the count of different network parameters \cite{2020Access}. M. Zolotukhin et al. proposed an intelligent defense system implemented as a reinforcement machine learning agent that processes current network state and takes a set of necessary actions in form of software-defined networking flows to redirect certain network traffic to virtual appliances \cite{netsoft2020}. A framework was also proposed which consists of a controller pool containing SD-IoT controllers, SD-IoT switches integrated with an IoT gateway, and IoT devices for software-defined Internet of Things based on the software-defined anything paradigm \cite{SDIoT2018}. In another study, J. Zheng et al. offered an easy and lightweight defense strategy against DDoS attacks against IoT devices in an SDN environment using the Markov Decision Process which optimal policies for handling network flows are specified to prevent DDoS attacks \cite{bigdata2018}.

\section{Network-aware AutoML Framework}
\label{sec:autoML}

\begin{figure}[htbp]
    \centering
    \includegraphics[width=0.47\textwidth]{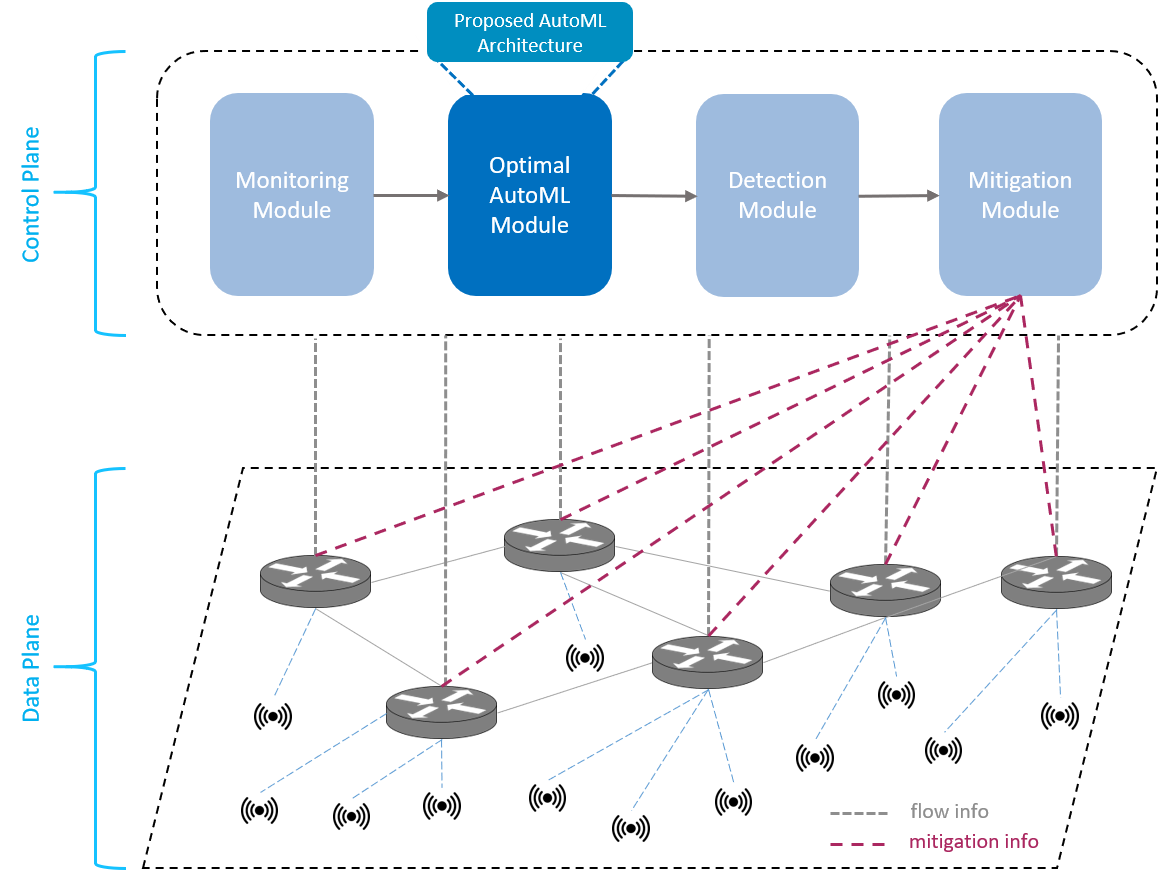}
    \caption{The targeted system architecture}
    \label{fig:architecture}
\end{figure}

The targeted system architecture in this study is shown in figure~\ref{fig:architecture}.
The proposed system works both in real-time and non-real-time. We have a buffer time for non-real-time and choose the ML algorithm which has the best performance for the network thanks to this buffer. We periodically pull data from the network, process them in non-real-time, and select the algorithm with the best performance for the current network situation, according to the result. Since we use a simple SDSN topology in this work, we have defined buffer time between 60 and 300 secs and pull the data from the network according to this range. When the buffer time comes, the optimal AutoML module decides the ideal algorithm for the real-time system.

\begin{figure}[htbp]
    \centering
    \includegraphics[width=0.52\textwidth]{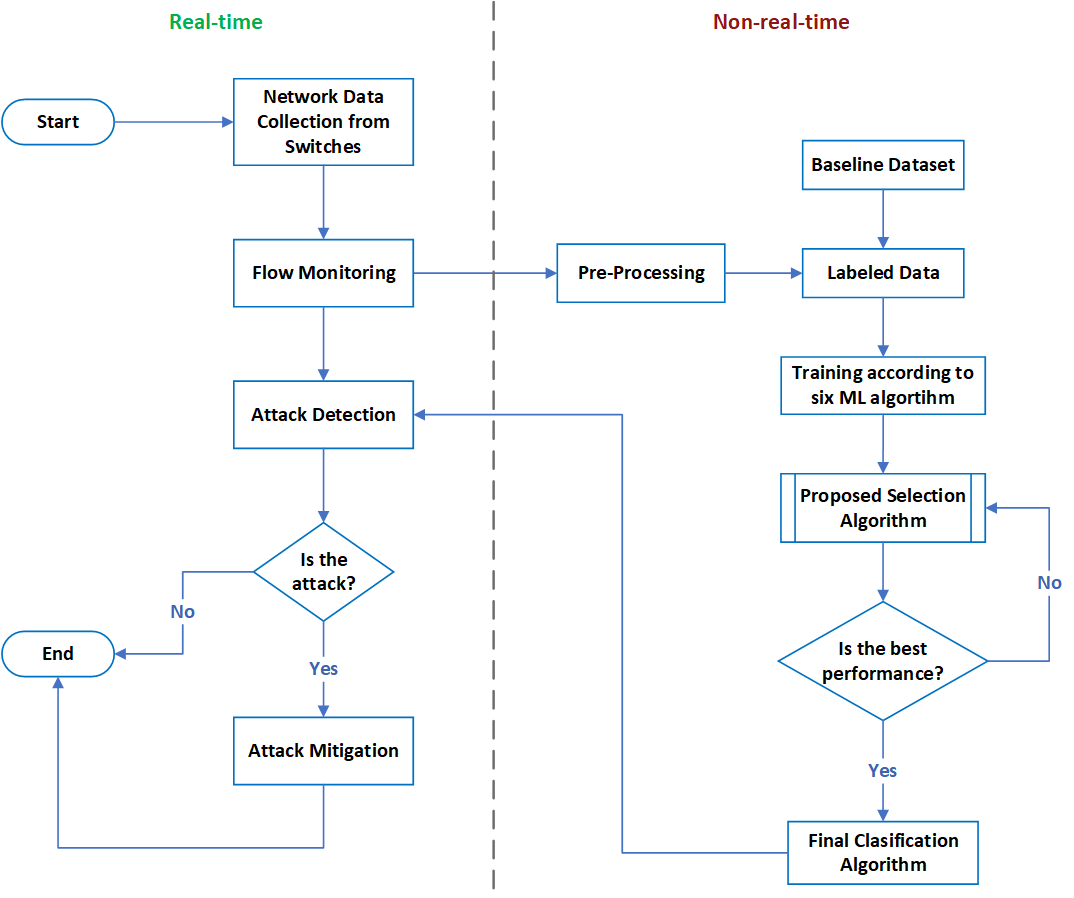}
    \caption{Flow Diagram of the proposed system}
    \label{fig:flow}
\end{figure}

The proposed system consists of four main modules:
\begin{itemize}
\item Monitoring
\item Optimal AutoML
\item Detection
\item Mitigation
\end{itemize}
Basically, these modules work as follows: The monitoring module receives the current data from the network and observes the flows, the optimal AutoML module finds the best ML algorithm for the network, the detection module identifies the attack flows using the ML algorithm, and the mitigation module drops the attack flows detected in the system. The flow diagram of the proposed system is shown in figure \ref{fig:flow}.
In addition, RYU 4.34 SDN Controller with Python 3.7.10, Mininet 2.3.0 as a simulator of virtual network, Ubuntu 16.04 operation system, D-ITG 2.8.1 as a traffic generator and Scikit-Learn 0.24.2 which is machine learning  library are used in our testbed.
\begin{figure*}[htbp]
    \centering
    \includegraphics[width=0.85\textwidth]{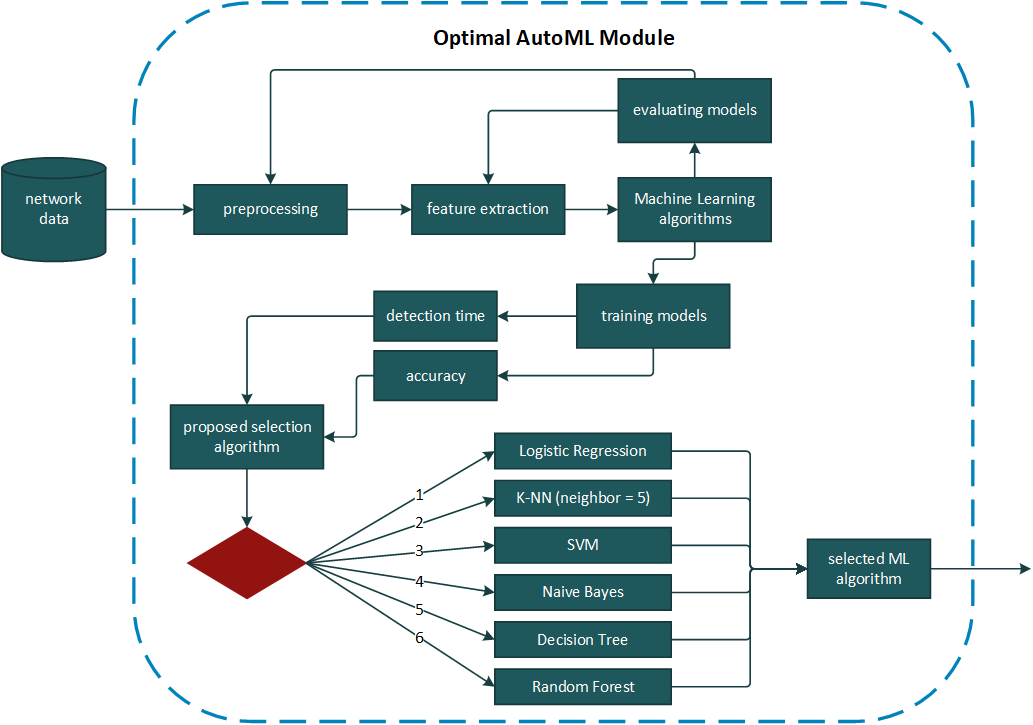}
    \caption{Architecture of optimal the AutoML Module}
    \label{fig:AutoML}
\end{figure*}
\subsection{Monitoring Module}
We use the OpenFlow protocol in our testbed. SDN only receives "PACKET IN" for the first packets that do not match in the flow table. It then creates a proper stream for later use, so SDN does not receive every packet on the network. Therefore, additional operations are required to know the current state of the network in SDN. It is possible to receive information about active flows using the OFPMP\_FLOW request from OpenFlow switches. In our system, the SDN controller creates a thread to periodically request all flow entry statistics from every switch in every second. The data that includes pckt\_count, pckt\_size, eth\_src, eth\_dst, etc.

\subsection{Optimal AutoML Module}
After monitoring traffic flows in the monitoring module, variable traffic payload, node  power  levels, and heterogeneous traffic speed parameters are received by the optimal AutoML module. The architecture of the AutoML Module is shown in figure~\ref{fig:AutoML}. As can be seen from system architecture, after taking network data by optimal AutoML module, machine learning algorithm which will be used by system decided according to the proposed selection algorithm.

\subsubsection{Proposed Selection Algorithm}
This algorithm considers detection time and accuracy. In the proposed selection algorithm, we use below optimization objective:
\begin{equation}
   arg\,max \:(\alpha_i A_i + \beta_i B_i), \:\: i \in [1, 6] 
\end{equation}
In equation 1, \emph{\(A_i\)} is accuracy of \emph{\(i_{th}\)} ML algorithm and \emph{\(B_i\)} is detection time of \emph{\(i_{th}\)} ML algorithm. We find the values \(\alpha\) and \(\beta\) separately for each ML algorithm. Then, these values are used to find the best performing ML algorithm according to the current network situation.

To calculate accuracy, we use the count of correct predictions in the related ML algorithm. In equation 2: \emph{y'} is predicted value, \emph{y} is real value, and \emph{n} is number of samples to calculate accuracy. 
\begin{equation}
   A_i(y,\:y') = \frac{1}{n} \: \sum_{j=0}^{n-1} \: {1}{(y'_j = y_j)}, \:\: i \in [1, 6] 
\end{equation}
\begin{equation}
    B_i = t^{end}_{i}\:-\:t^{start}_{i}, \:\: i \in [1, 6] 
\end{equation}
Equation 2 exhibits the accuracy calculation of \emph{\(i_{th}\)} ML algorithm. Since we design the proposed system for the network-constrained environments, the detection time is another important metric. We calculate the detection time using finish time \emph{(\(t_{exit}\))} and starting time \emph{(\(t_{start}\))} of \emph{\(i_{th}\)} ML algorithm. 
Equation 3 represents the calculation of the detection time.

\subsection{Detection Module}

After selecting an ML algorithm that has the best performance by the optimal AutoML module, the data are trained which pull from the network. After all processes, data are labeled as \emph{Type 1 (DDoS)} or \emph{Type 0 (normal)} appropriately according to our baseline dataset. The creation of our baseline dataset is explained in section~\ref{sec:perfEval}. When the proposed system detects a DDoS attack, an alert is sent to the system.

\subsection{Mitigation Module}
When the flow has a DDoS attack is detected, the flow is deleted by sending DEL command to the switches. This process provides switches that send "PACKET IN" to the SDN controller when the next packet arrives from that source to that destination. Since the SDN controller is deleted the flow due to a possible DDoS attack, it does not immediately create a new flow for these "PACKET IN"s, instead, it waits for a while and then creates a new flow. This means that during a DDoS attack, packets are dropped, not delivered and the network continues to operate normally.

\section{Performance Evaluation}
\label{sec:perfEval}

\begin{table}[h]
\caption{Simulation Parameters}
\label{table:simulation}
\centering
\begin{tabular}{|c|c|c|} 
\hline
\multicolumn{2}{|c|}{\textbf{Parameters}} & \textbf{Values}\\ 
\hline

\multirow{3}{*}[0.3em]{\rotatebox[origin=c]{90}{ \shortstack[c]{ Variable \\ Payload }
 }} & small size & {100-999} byte \\
 \cline{2-3}
 & medium size & {1000-9999} byte\\
 \cline{2-3}
 & large size & {10000-99999} byte\\
\hline

\multirow{3}{*}[0.2em]{\rotatebox[origin=c]{90}{ \shortstack[c]{ 
Traffic \\ Speed}
 }} & low speed & {1-100} packet per second (pps)\\
 \cline{2-3}
 & moderate speed & {101-1000} pps\\
 \cline{2-3}
 & fast speed & {1001-10000} pps\\
 \hline
 
\multicolumn{2}{|c|}{Number of Nodes} & { \begin{tabular}[c]{@{}l@{}}100, 200, 300 , 400, 500, \\600, 700, 800, 900, 1000\end{tabular}}\\ 
\hline

\end{tabular}
\end{table}

\begin{figure}[htbp]
    \centering
    \includegraphics[width=0.45\textwidth]{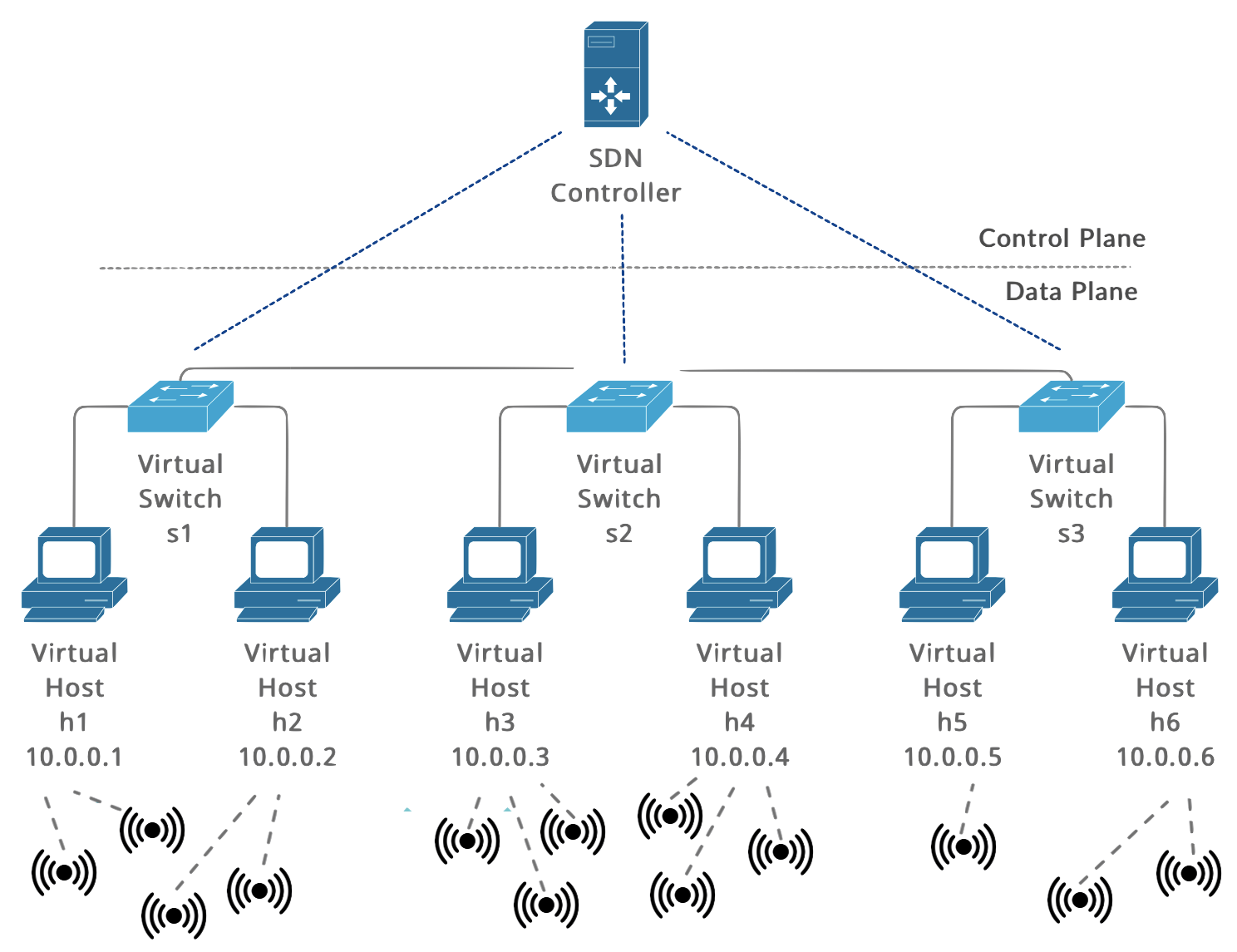}
    \caption{Mininet Topology}
    \label{fig:top}
\end{figure}

Simulation parameters are presented in table~\ref{table:simulation}.
We create a simple SDSN topology to create baseline dataset. Our baseline dataset consist of the nine different datasets and these datasets are created according to variable traffic payload and heterogeneous traffic speed criteria. To create a baseline dataset, a topology is used as in figure~\ref{fig:top} in Mininet. D-ITG tool is used to generate traffic both normal traffic and DDoS attack using the UDP packets. While the traffic is running, all statistics from the monitoring module are saved in a ".csv" file. Data are labeled as Type 1 (DDoS) or Type 0 (normal) appropriately, to be used as the dataset.

\begin{figure}[htbp]
    \centering
    \includegraphics[width=0.5\textwidth]{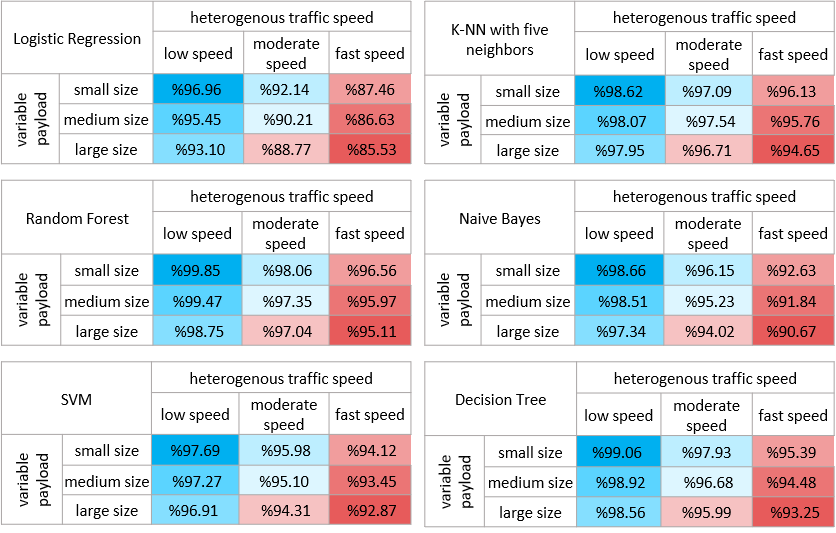}
    \caption{The test accuracy of ML algorithms for datasets}
    \label{fig:accuracy}
\end{figure}
\begin{figure}[htbp]
    \centering
    \includegraphics[width=0.5\textwidth]{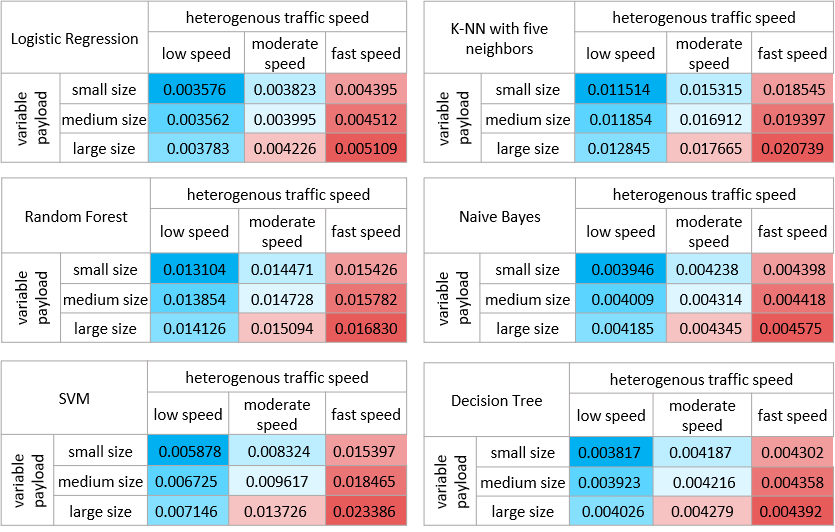}
    \caption{The detection time of ML algorithms for datasets}
    \label{fig:detectiontime}
\end{figure}
We test the ML algorithms using nine different datasets to investigate accuracy and detection time of the algorithms. 
Figure~\ref{fig:accuracy} and figure~\ref{fig:detectiontime} showcase the accuracy and detection time results of the six ML algorithms respectively. 
The test results point out that although the Random Forest algorithm has the best accuracy, the Decision Tree algorithm offers the best performance compared to the others.

\begin{figure}[htbp]
    \centering
    \includegraphics[width=0.45\textwidth]{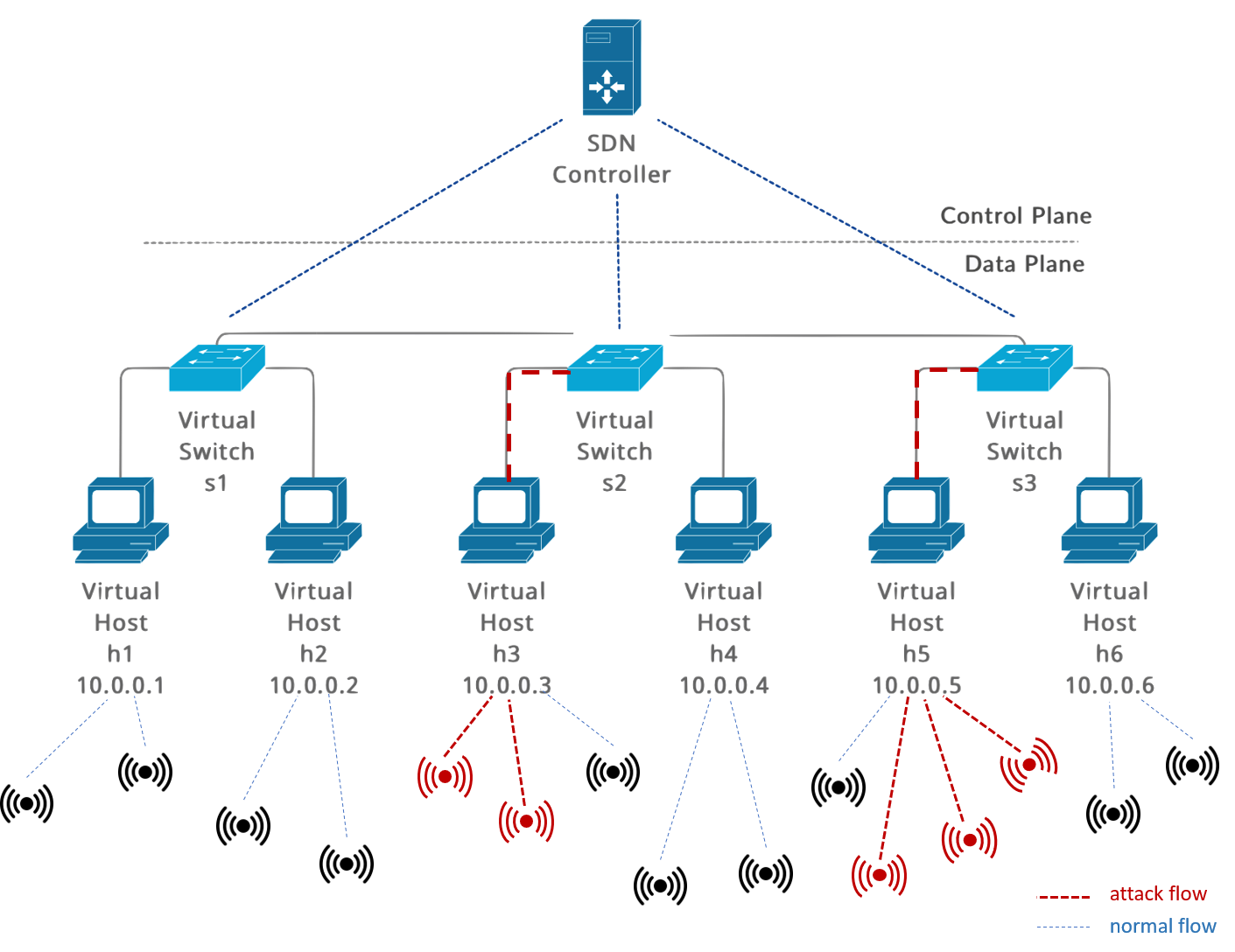}
    \caption{Test Topology}
    \label{fig:attack}
\end{figure}
\begin{figure}[htbp]
    \centering
    \includegraphics[width=0.44\textwidth]{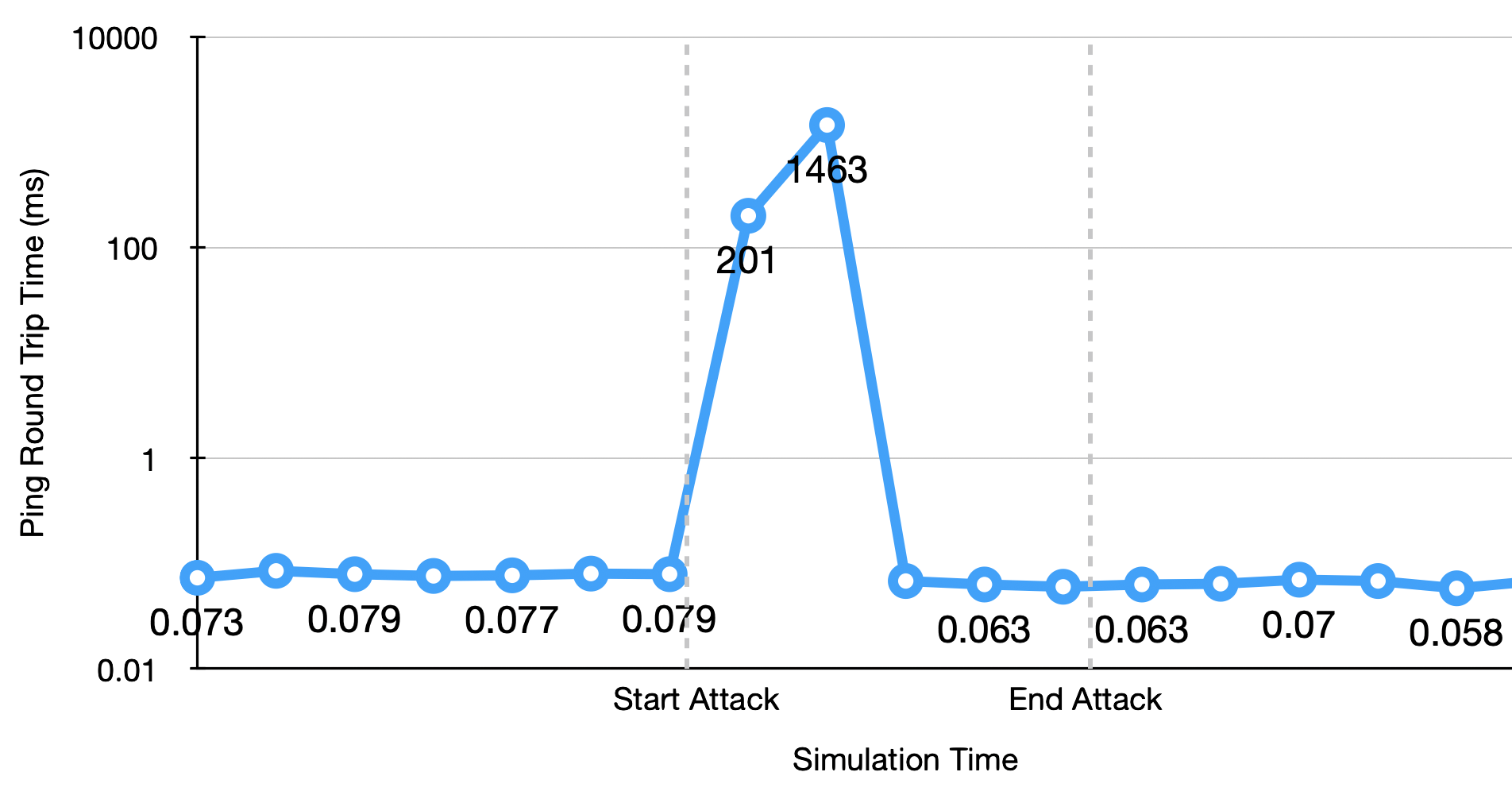}
    \caption{Round Trip Time for the proposed system}
    \label{fig:RTT}
\end{figure}
\begin{figure}[htbp]
    \centering
    \includegraphics[width=0.44\textwidth]{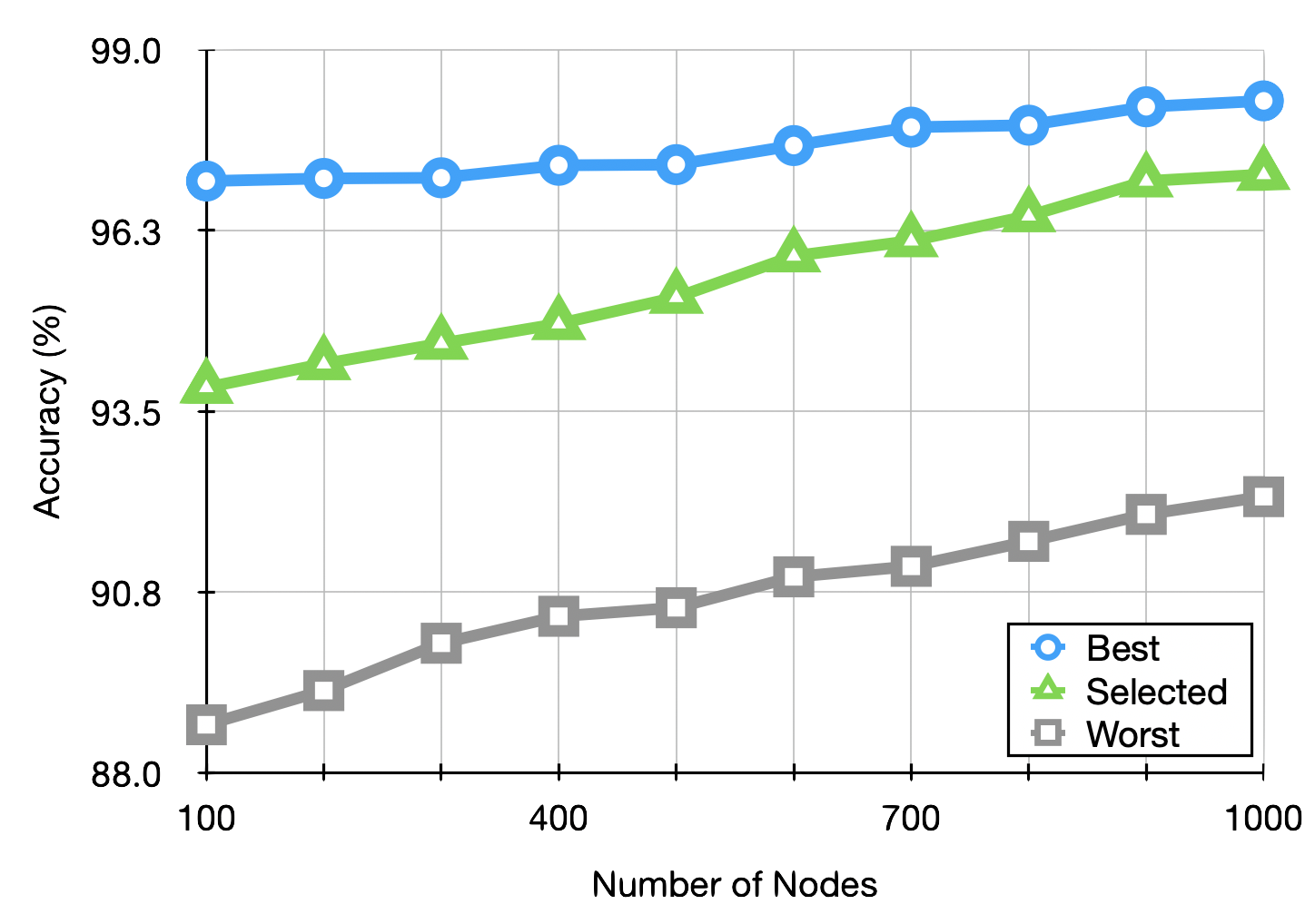}
    \caption{Accuracy for the algorithms}
    \label{fig:acc}
\end{figure}

After this investigation, we test our system design using the topology in figure~\ref{fig:attack}. 
The target host is Host 1 (10.0.0.1), the attacker attacks on Host 3 (10.0.0.3) and Host 5 (10.0.0.5) through sensors, and Host 6 (10.0.0.6) is used to ping Host 1 to measure the round trip time.
\begin{figure}[htbp]
    \centering
    \includegraphics[width=0.44\textwidth]{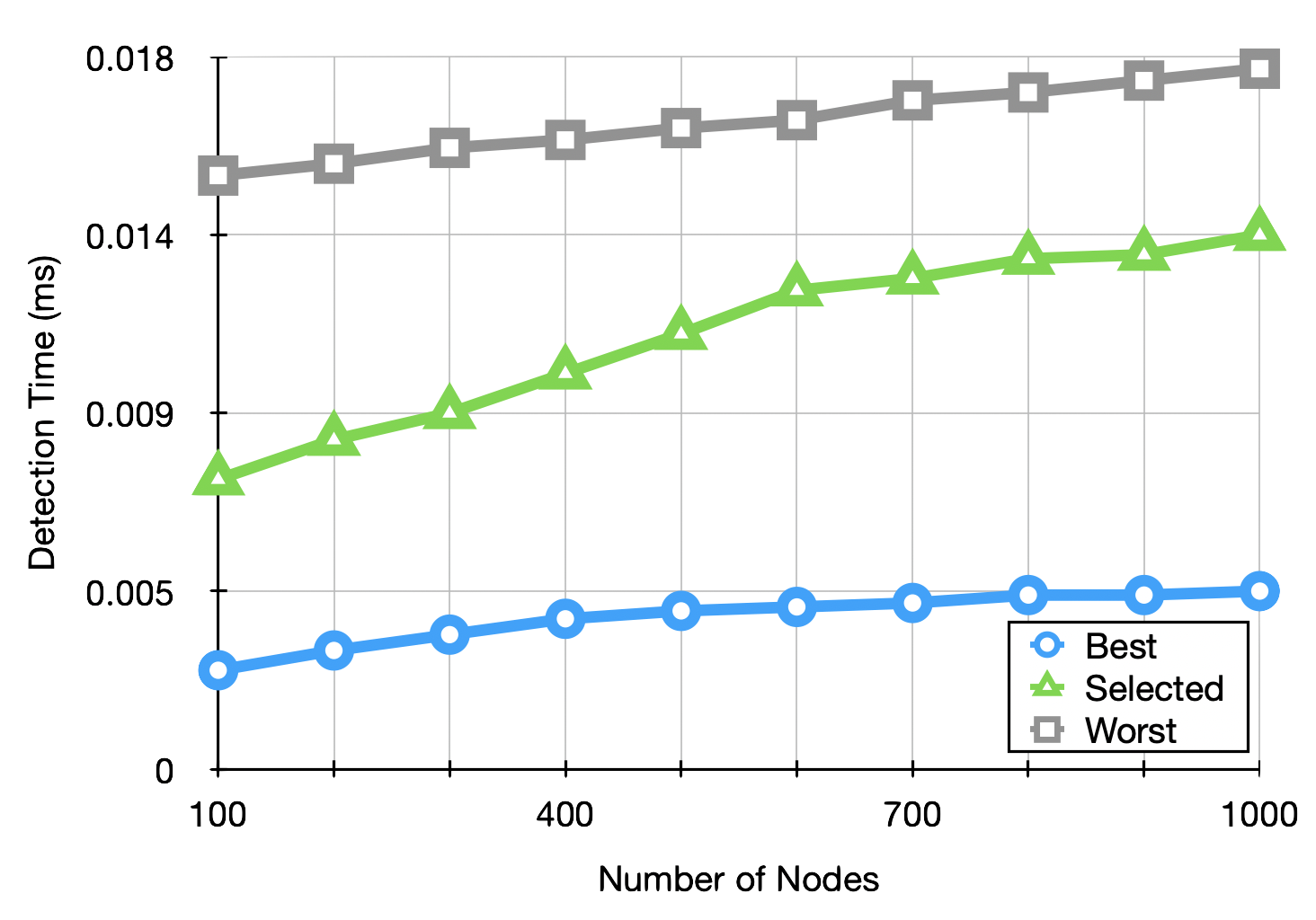}
    \caption{Detection time for the algorithms}
    \label{fig:det}
\end{figure}
When we pull the data from the network and the monitoring module displays the flows. Then, the optimal AutoML module decides on the best machine learning algorithm in non-real-time. The detection module checks the flows to detect DDoS attacks. When the proposed system detects a DDoS attack, an alert is sent to the system. After that, the mitigation module mitigates the flow that has the DDoS attack.

When we implement the proposed system to our testbed, our system result has a better round trip time than in section~\ref{sec:preliminary} and figure~\ref{fig:RTT} shows round trip time graph of the proposed system.
Since we choose the algorithm with the best performance in network awareness, the chosen algorithm does not have the best accuracy and the best detection time. The graph in figure~\ref{fig:acc} shows the accuracy results of the best, worst, and selected algorithms of the system. Detection time results are shown in figure~\ref{fig:det}.

\section{Conclusion}
\label{sec:conc}
As a conclusion, we designed and implemented an automated machine learning framework for DDoS attack in software-defined sensor networks. 
The proposed framework select most suitable ML algorithm for the sensor network ensuring computational capacity of the model can be sustained by the network. 
Finally, we showed that our framework successfully mitigates disruptions in the network while the packets are delivered with additional delays. 
In future work, we will expand our framework using a more complex topology and set the buffer time interval between 1 and 60 seconds to test our framework by creating a real-life software-defined sensor network scenario.

\section*{Acknowledgment}
Yagmur Yigit would like to thank the DeepMind Scholarship programme for their support.

\bibliographystyle{IEEEtran}

\begin{thebibliography}{10}
\providecommand{\url}[1]{#1}
\csname url@samestyle\endcsname
\providecommand{\newblock}{\relax}
\providecommand{\bibinfo}[2]{#2}
\providecommand{\BIBentrySTDinterwordspacing}{\spaceskip=0pt\relax}
\providecommand{\BIBentryALTinterwordstretchfactor}{4}
\providecommand{\BIBentryALTinterwordspacing}{\spaceskip=\fontdimen2\font plus
\BIBentryALTinterwordstretchfactor\fontdimen3\font minus
  \fontdimen4\font\relax}
\providecommand{\BIBforeignlanguage}[2]{{%
\expandafter\ifx\csname l@#1\endcsname\relax
\typeout{** WARNING: IEEEtran.bst: No hyphenation pattern has been}%
\typeout{** loaded for the language `#1'. Using the pattern for}%
\typeout{** the default language instead.}%
\else
\language=\csname l@#1\endcsname
\fi
#2}}
\providecommand{\BIBdecl}{\relax}
\BIBdecl

\bibitem{iot2019}
K.~R. Sollins, ``Iot big data security and privacy versus innovation,''
  \emph{IEEE Internet of Things Journal}, vol.~6, no.~2, pp. 1628--1635, 2019.

\bibitem{survey44}
W.~Rafique, L.~Qi, I.~Yaqoob, M.~Imran, R.~U. Rasool, and W.~Dou,
  ``Complementing iot services through software defined networking and edge
  computing: A comprehensive survey,'' \emph{IEEE Communications Surveys
  Tutorials}, vol.~22, no.~3, pp. 1761--1804, 2020.

\bibitem{IoT2017}
C.~Kolias, G.~Kambourakis, A.~Stavrou, and J.~Voas, ``Ddos in the iot: Mirai
  and other botnets,'' \emph{Computer}, vol.~50, no.~7, pp. 80--84, 2017.

\bibitem{transSDN2018}
J.~Zheng, Q.~Li, G.~Gu, J.~Cao, D.~K.~Y. Yau, and J.~Wu, ``Realtime ddos
  defense using cots sdn switches via adaptive correlation analysis,''
  \emph{IEEE Transactions on Information Forensics and Security}, vol.~13,
  no.~7, pp. 1838--1853, 2018.

\bibitem{challengeSDSN}
H.~I. Kobo, A.~M. Abu-Mahfouz, and G.~P. Hancke, ``A survey on software-defined
  wireless sensor networks: Challenges and design requirements,'' \emph{IEEE
  Access}, vol.~5, pp. 1872--1899, 2017.

\bibitem{trans20intro}
Q.~Li, X.~He, M.~Xu, Y.~Jiang, and L.~Wang, ``Unified middlebox model design
  and deployment with dynamic resources,'' \emph{IEEE Transactions on Network
  and Service Management}, vol.~15, no.~3, pp. 1035--1048, 2018.

\bibitem{africon2017}
T.~Kgogo, B.~Isong, and A.~M. Abu-Mahfouz, ``Software defined wireless sensor
  networks security challenges,'' in \emph{2017 IEEE AFRICON}, 2017, pp.
  1508--1513.

\bibitem{19MLAuto}
Z.~Li, H.~Guo, W.~M. Wang, Y.~Guan, A.~V. Barenji, G.~Q. Huang, K.~S. McFall,
  and X.~Chen, ``A blockchain and automl approach for open and automated
  customer service,'' \emph{IEEE Transactions on Industrial Informatics},
  vol.~15, no.~6, pp. 3642--3651, 2019.

\bibitem{Ariman}
M.~Ariman, G.~Seçinti, M.~Erel, and B.~Canberk, ``Software defined wireless
  network testbed using raspberry pi of switches with routing add-on,'' in
  \emph{2015 IEEE Conference on Network Function Virtualization and Software
  Defined Network (NFV-SDN)}, 2015, pp. 20--21.

\bibitem{det2}
B.~Nugraha and R.~N. Murthy, ``Deep learning-based slow ddos attack detection
  in sdn-based networks,'' in \emph{2020 IEEE Conference on Network Function
  Virtualization and Software Defined Networks (NFV-SDN)}, 2020, pp. 51--56.

\bibitem{detmit2}
K.~Hong, Y.~Kim, H.~Choi, and J.~Park, ``Sdn-assisted slow http ddos attack
  defense method,'' \emph{IEEE Communications Letters}, vol.~22, no.~4, pp.
  688--691, 2018.

\bibitem{20Access}
L.~Tan, Y.~Pan, J.~Wu, J.~Zhou, H.~Jiang, and Y.~Deng, ``A new framework for
  ddos attack detection and defense in sdn environment,'' \emph{IEEE Access},
  vol.~8, pp. 161\,908--161\,919, 2020.

\bibitem{19ICC}
Y.~Chen, J.~Pei, and D.~Li, ``Detpro: A high-efficiency and low-latency system
  against ddos attacks in sdn based on decision tree,'' in \emph{ICC 2019 -
  2019 IEEE International Conference on Communications (ICC)}, 2019, pp. 1--6.

\bibitem{20AcKNN}
S.~Dong and M.~Sarem, ``Ddos attack detection method based on improved knn with
  the degree of ddos attack in software-defined networks,'' \emph{IEEE Access},
  vol.~8, pp. 5039--5048, 2020.

\bibitem{19ICTC}
N.~N. Tuan, P.~H. Hung, N.~D. Nghia, N.~Van~Tho, T.~V. Phan, and N.~H. Thanh,
  ``A robust tcp-syn flood mitigation scheme using machine learning based on
  sdn,'' in \emph{2019 International Conference on Information and
  Communication Technology Convergence (ICTC)}, 2019, pp. 363--368.

\bibitem{20LowRate}
J.~A. Pérez-Díaz, I.~A. Valdovinos, K.-K.~R. Choo, and D.~Zhu, ``A flexible
  sdn-based architecture for identifying and mitigating low-rate ddos attacks
  using machine learning,'' \emph{IEEE Access}, vol.~8, pp. 155\,859--155\,872,
  2020.

\bibitem{19Congress}
O.~Rahman, M.~A.~G. Quraishi, and C.-H. Lung, ``Ddos attacks detection and
  mitigation in sdn using machine learning,'' in \emph{2019 IEEE World Congress
  on Services (SERVICES)}, vol. 2642-939X, 2019, pp. 184--189.

\bibitem{kubra}
K.~Duran, B.~Karanlik, and B.~Canberk, ``Graph theoretical approach for
  automated ip lifecycle management in telco networks,'' \emph{International
  Journal of Network Management}, 09 2020.

\bibitem{20Globe}
A.~Ahmad, E.~Harjula, M.~Ylianttila, and I.~Ahmad, ``Evaluation of machine
  learning techniques for security in sdn,'' in \emph{2020 IEEE Globecom
  Workshops (GC Wkshps}, 2020, pp. 1--6.

\bibitem{9369564}
K.~Duran, B.~Karanlik, and B.~Canberk, ``Ai-driven partial topology discovery
  algorithm for broadband networks,'' in \emph{2021 IEEE 18th Annual Consumer
  Communications Networking Conference (CCNC)}, 2021, pp. 1--6.

\bibitem{2019wiley}
R.~Santos, D.~Silva, W.~Santo, A.~Ribeiro, and E.~Ordonez, ``Machine learning
  algorithms to detect ddos attacks in sdn,'' \emph{Concurrency and
  Computation: Practice and Experience}, vol.~32, p. e5402, 06 2019.

\bibitem{Ak}
E.~Ak and B.~Canberk, ``Fsc: Two-scale ai-driven fair sensitivity control for
  802.11ax networks,'' in \emph{GLOBECOM 2020 - 2020 IEEE Global Communications
  Conference}, 2020, pp. 1--6.

\bibitem{ICC2020}
G.~A.~N. Segura, S.~Skaperas, A.~Chorti, L.~Mamatas, and C.~B. Margi, ``Denial
  of service attacks detection in software-defined wireless sensor networks,''
  in \emph{2020 IEEE International Conference on Communications Workshops (ICC
  Workshops)}, 2020, pp. 1--7.

\bibitem{2020Access}
J.~Bhayo, S.~Hameed, and S.~A. Shah, ``An efficient counter-based ddos attack
  detection framework leveraging software-defined iot (sd-iot),'' \emph{IEEE
  Access}, vol.~8, pp. 221\,612--221\,631, 2020.

\bibitem{netsoft2020}
M.~Zolotukhin, S.~Kumar, and T.~Hämäläinen, ``Reinforcement learning for
  attack mitigation in sdn-enabled networks,'' in \emph{2020 6th IEEE
  Conference on Network Softwarization (NetSoft)}, 2020, pp. 282--286.

\bibitem{SDIoT2018}
D.~Yin, L.~Zhang, and K.~Yang, ``A ddos attack detection and mitigation with
  software-defined internet of things framework,'' \emph{IEEE Access}, vol.~6,
  pp. 24\,694--24\,705, 2018.

\bibitem{bigdata2018}
J.~Zheng and A.~S. Namin, ``Defending sdn-based iot networks against ddos
  attacks using markov decision process,'' in \emph{2018 IEEE International
  Conference on Big Data (Big Data)}, 2018, pp. 4589--4592.

\end{thebibliography}


\end{document}